\documentclass[10pt]{article}
\usepackage{fullpage}
\usepackage{amsmath}
\usepackage{amssymb}
\usepackage[dvips]{epsfig}
\usepackage{color}

\def\##1{\underline{#1}}
\def\=#1{\underline{\underline{#1}}}

\def\+#1{\underline{\bf #1}}
\def\*#1{\underline{\underline{\bf #1}}}

\def\r#1{(\ref{#1})}
\def\l#1{\label{#1}}
\def\c#1{\cite{#1}}

\def\le{\left(}
\def\ri{\right)}
\def\les{\left[}
\def\ris{\right]}
\def\lec{\left\{}
\def\ric{\right\}}

\def\.{\mbox{ \tiny{$^\bullet$} }}

\def\epso{\epsilon_{\scriptscriptstyle 0}}

\def\muo{\mu_{\scriptscriptstyle 0}}

\def\ko{k_{\scriptscriptstyle 0}}

\def\Eo{\#E_{\, \scriptscriptstyle 0}}
\def\Ho{\#H_{\, \scriptscriptstyle 0}}

\begin{document}

\begin{center}

{\bf {\LARGE Ray trajectories for a spinning cosmic string and a
manifestation of self--cloaking}}

\vspace{10mm} \large

 Tom H. Anderson\footnote{E--mail: T.H.Anderson@sms.ed.ac.uk}\\
{\em School of Mathematics and
   Maxwell Institute for Mathematical Sciences\\
University of Edinburgh, Edinburgh EH9 3JZ, UK}\\
 \vspace{3mm}
 Tom G. Mackay\footnote{E--mail: T.Mackay@ed.ac.uk}\\
{\em School of Mathematics and
   Maxwell Institute for Mathematical Sciences\\
University of Edinburgh, Edinburgh EH9 3JZ, UK}\\
and\\
 {\em NanoMM~---~Nanoengineered Metamaterials Group\\ Department of Engineering Science and Mechanics\\
Pennsylvania State University, University Park, PA 16802--6812,
USA}\\
 \vspace{3mm}
 Akhlesh  Lakhtakia\footnote{E--mail: akhlesh@psu.edu}\\
 {\em NanoMM~---~Nanoengineered Metamaterials Group\\ Department of Engineering Science and Mechanics\\
Pennsylvania State University, University Park, PA 16802--6812, USA}

\end{center}

\vspace{4mm}

\normalsize

\begin{abstract}

A study of ray trajectories was undertaken  for the Tamm medium
which represents the spacetime of a cosmic spinning string, under
the geometric-optics approximation. Our numerical studies revealed
that: (i) rays never cross the string's boundary; (ii) the Tamm
medium supports evanescent waves in regions of phase space that
correspond to those regions of the string's  spacetime which could support
closed timelike curves; and (iii) a spinning string
can be slightly visible while a
  non--spinning string is almost perfectly invisible.
\end{abstract}


\vspace{5mm} \noindent  {\bf Keywords:} cosmic spinning string, Tamm
medium, invisibility, metamaterial, ray tracing

\section{Introduction}

Parallels have been established
 between   exotic optical phenomenons associated with certain
metamaterials and those associated with curved spacetimes
\c{de_Sitter_metamaterial,Black_hole_metamaterial,css}.
 For
example, negative-phase-velocity propagation of light~---~a notable
property supported by certain negatively refracting metamaterials
\c{ML_PRB}~---~is  associated with various spacetime metrics in
a noncovariant formalism \c{MLS_NJP_Kerr,Kom,Sharif,Kerr_Newman,Hossain}.
 This parallelism is perhaps not surprising given that
light propagation in vacuum subjected to a gravitational field is
formally equivalent to light propagation in a nonhomogeneous
bianisotropic medium~---~called a Tamm medium \c{Tamm}~---~in flat
spacetime \c{Skrotskii,Plebanski,SS}. The practical realization of
Tamm mediums is edging ever closer due to rapid
 advances in the science of nanostructured materials, especially those relating to
  metamaterials supporting negative refraction or
 cloaking applications \c{Huebner,He}.

The constitutive relations for the Tamm
medium representing a spinning cosmic string were recently established \c{css}. Cosmic strings
are topological defects associated with phase transitions which may
represent traces of the very early universe, as is comprehensively
described elsewhere
 \c{Hindmarsh,Vilenkin}. In the following sections we report on the
 trajectories of light rays in the Tamm medium representing the spacetime metric of a cosmic
 spinning string, in the geometric-optics regime. In particular, we
 report on a cloaking phenomenon~---~analogous to that associated
 with certain metamaterials \c{Greenleaf}~---~which arises when the cosmic string's angular momentum is set to zero,
 thereby rendering it almost perfectly invisible to optical probes.

\section{Quasi--planewave analysis}

We consider   a cosmic spinning string that is aligned parallel to the $z$
axis. Let $\rho = \sqrt{x^2 + y^2}$. The `ballpoint pen' model
\c{Jensen_Soleng} is adopted wherein
  the spacetime  metric describing the string's interior region
 $\rho < \rho_s$ is matched smoothly at $\rho=\rho_s$ to the spacetime metric
describing the string's exterior region $\rho > \rho_s$. The Tamm
medium representing this cosmic spinning string is characterized by
the constitutive relations \c{css}
\begin{equation}
\left. \begin{array}{l} \#D (\#r,t) = \epso \=\gamma (x, y) \. \#E
(\#r,t) - \sqrt{\epso \muo} \; \#\Gamma (x, y) \times \#H
(\#r,t)\vspace{4pt}
 \\
\#B (\#r,t) = \muo \=\gamma (x, y) \. \#H (\#r,t) + \sqrt{\epso \muo} \;
\#\Gamma (x, y) \times \#E (\#r,t)
\end{array}
\right\}
 \l{CR}
\end{equation}
in SI units. The scalar constants $\epso$ and $\muo$ denote the
permittivity and permeability of  vacuum in the absence of a
gravitational field, while the 3$\times$3 dyadic
\begin{equation}
\=\gamma (x, y) = \frac{1}{\rho} \le
\begin{array}{ccc}
\displaystyle{\frac{A^2 x^2 + \rho^2 y^2}{A \rho^2}} &
\displaystyle{\frac{\le A^2 - \rho^2 \ri x y }{A \rho^2}} & 0
\\ \\
\displaystyle{\frac{\le A^2 - \rho^2 \ri x y }{A \rho^2}} &
\displaystyle{\frac{A^2 y^2 + \rho^2 x^2}{A \rho^2}} &0 \\ \\ 0 & 0
& A
\end{array}
 \ri \l{matrix_gamma}
\end{equation}
and the 3--vector
\begin{equation}
\#\Gamma (x, y)  = \frac{M}{\rho^2}  \le -y, \, x, \, 0 \ri
\l{vector_gamma}.
\end{equation}
Herein the scalar quantities
\begin{equation}
A =  \left\{
\begin{array}{lr}
\displaystyle{\frac{1}{\sqrt{\lambda}} \, \sin \le \rho
\sqrt{\lambda}
 \ri,} & \rho \leq \rho_s \vspace{6pt}\\ \displaystyle{\cos \le \rho_s
\sqrt{\lambda} \ri \les \rho + \rho_s \le \frac{\tan \le \rho_s
\sqrt{\lambda} \ri }{\rho_s \sqrt{\lambda}} -1 \ri \ris,} & \rho >
\rho_s
\end{array}
\right.
\end{equation}
and
\begin{equation}
M =  \left\{
\begin{array}{lr}
\displaystyle{2 \alpha \les \le \rho - \rho_s \ri \cos \le \rho
\sqrt{\lambda}
 \ri  - \frac{1}{\sqrt{\lambda}} \, \sin \le
\rho \sqrt{\lambda}  \ri + \rho_s \ris,} & \rho \leq \rho_s \vspace{6pt}\\
 \displaystyle{ 2 \alpha   \les  -
\frac{1}{\sqrt{\lambda}} \, \sin \le \rho_s \sqrt{\lambda}  \ri +
\rho_s \ris} & \rho
> \rho_s
\end{array}
\right.
\end{equation}
are expressed in terms of the real--valued constants $\lambda$ and
$\alpha$, which are related to the energy density and angular
momentum respectively. As we discussed previously  \c{css}, the
choice $\lambda = \le \pi / \rho_s \ri^2$ ensures that the
constitutive parameters of the  Tamm medium coincide with those of
vacuum (in flat spacetime) in the limits $\rho \to 0$ and $\infty$.
In keeping with our earlier study \c{css},  we set $\alpha = 1$ for
our numerical studies of a spinning string, and $\alpha = 0$ for a
non--spinning string.

Consistently with the geometric-optics approximation,  we consider
quasi--planewave  electric and magnetic fields of the form  \c{vanB}
\begin{equation}
\left.
\begin{array}{l}
 \#E (\#r,t) = {\rm Re} \lec \,\Eo(\#r) \exp \les i \le  \ko \#k\cdot \#r  - \omega
t \ri \ris\ric \vspace{6pt}
\\   \#H (\#r,t) = {\rm Re} \lec \, \Ho (\#r) \exp \les i \le  \ko \#k\cdot \#r - \omega t
\ri \ris\ric
\end{array}
\right\},
  \l{qpw}
\end{equation}
where $\Eo (\#r)$ and $\Ho (\#r)$ are complex--valued amplitudes,
$\omega$ is the angular frequency, and the vacuum wavenumber $\ko =
\omega \sqrt{\epso \muo}$. The relative wavevector $\#k$ in the
quasi--planewave representation \r{qpw} is a function of  $\#r$ but
for later convenience we omit explicit reference to $\#r$.
 Combining eqs.~\r{CR} and \r{qpw} with  the source--free Maxwell curl
postulates yields
\begin{eqnarray}
&& \les \nabla \le \#k\cdot \#r \ri - \#\Gamma (x,y) \ris  \times
\Eo (\#r) - \sqrt{\frac{\muo}{\epso}} \, \=\gamma (x, y) \cdot \Ho
(\#r)
= - \frac{1}{i \ko} \, \nabla \times \Eo (\#r), \l{m1}\\
&& \les \nabla \le \#k\cdot \#r \ri  - \#\Gamma (x,y) \ris  \times
\Ho (\#r) + \sqrt{\frac{\epso}{\muo}} \, \=\gamma (x, y) \cdot \Eo
(\#r) = - \frac{1}{i \ko} \, \nabla \times \Ho (\#r) \l{m2}.
\end{eqnarray}
In the geometric-optics regime, the constitutive parameters are assumed
to vary very slowly over the distance of a wavelength. Accordingly,
the right sides of eqs. \r{m1} and \r{m2} are set to zero, and  $
\nabla \le \#k\cdot \#r \ri \approx \#k$. Thus, eqs.~\r{m1} and \r{m2}
simplify to \c{LM1}
\begin{equation}
\lec \les \det \=\gamma (x, y) - \#p   \cdot \=\gamma (x, y ) \cdot
 \#p   \,\ris \=I + \#p  \, \#p  \cdot \=\gamma (x, y) \ric \cdot \Eo (\#r) =
 \#0 \,, \l{e1}
\end{equation}
where $\#p  =  \#k  - \#\Gamma (x, y)$ and $\=I$ denotes the
identity 3$\times$3 dyadic. The  dyadic enclosed in  braces on the
left side of eq. \r{e1} is required to  be nonsingular, in order for
nontrivial solutions to exist. This leads to the dispersion relation
\c{LM1}
\begin{equation}
\mathcal{H}  \equiv  \det \=\gamma (x, y) - \#p   \cdot \=\gamma (x,
y ) \cdot
 \#p  = 0, \l{disp}
\end{equation}
 from which the  magnitude $k$ of the relative wavevector $\#k$ may be deduced. In fact, the
 $k$--roots of the dispersion relation \r{disp}~---~corresponding
to  $\#k = k$ $ ( \sin \theta \,\cos \phi, $
$\, \sin \theta \,\sin \phi, $ $\, \cos \theta )$ at $\#r = \le \rho
\cos \sigma, \, \rho \sin \sigma, z \ri$~---~may be expressed as
\begin{equation}
k = \frac{-b \pm \sqrt{b^2 - 4 a c }}{2 a}, \l{kroot}
\end{equation}
where the coefficients
\begin{eqnarray}
a &=&   A^2 \cos^2 \theta + \les A^2 \cos^2 \le \sigma - \phi \ri +
\rho^2 \sin^2  \le \sigma - \phi \ri \ris \sin^2 \theta,\\
b &=& 2 M \rho \sin \theta \, \sin \le \sigma - \phi \ri, \\
c &= & M^2 - A^2.
\end{eqnarray}
Although eq.~\r{disp} is   a quadratic equation in $k$, it yields only one
independent root, in consonance with vacuum being unirefringent
\c{LMcpl}.

\begin{figure}[!ht]
\centering \psfull \epsfig{file=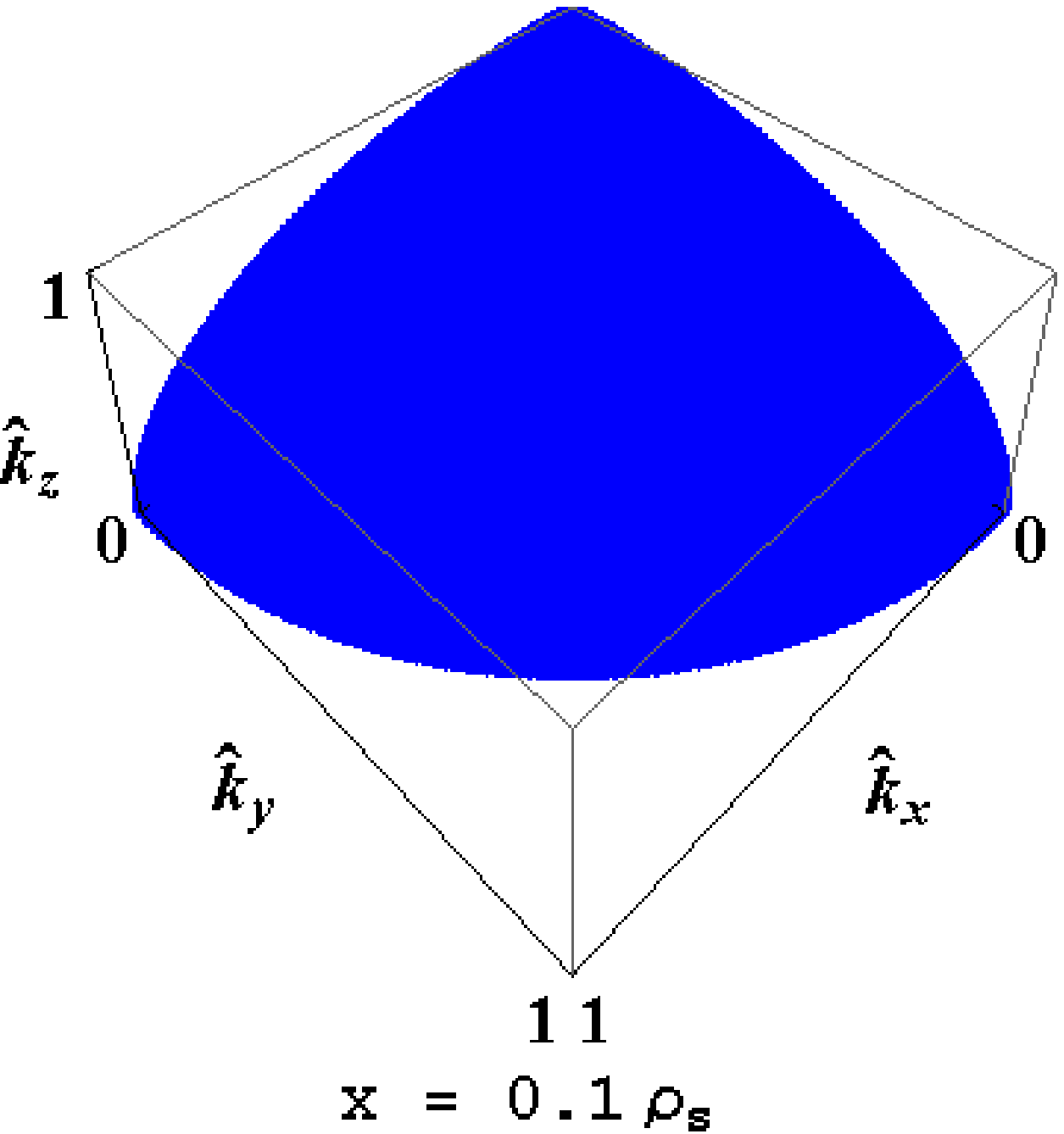,width=1.80in}
\epsfig{file=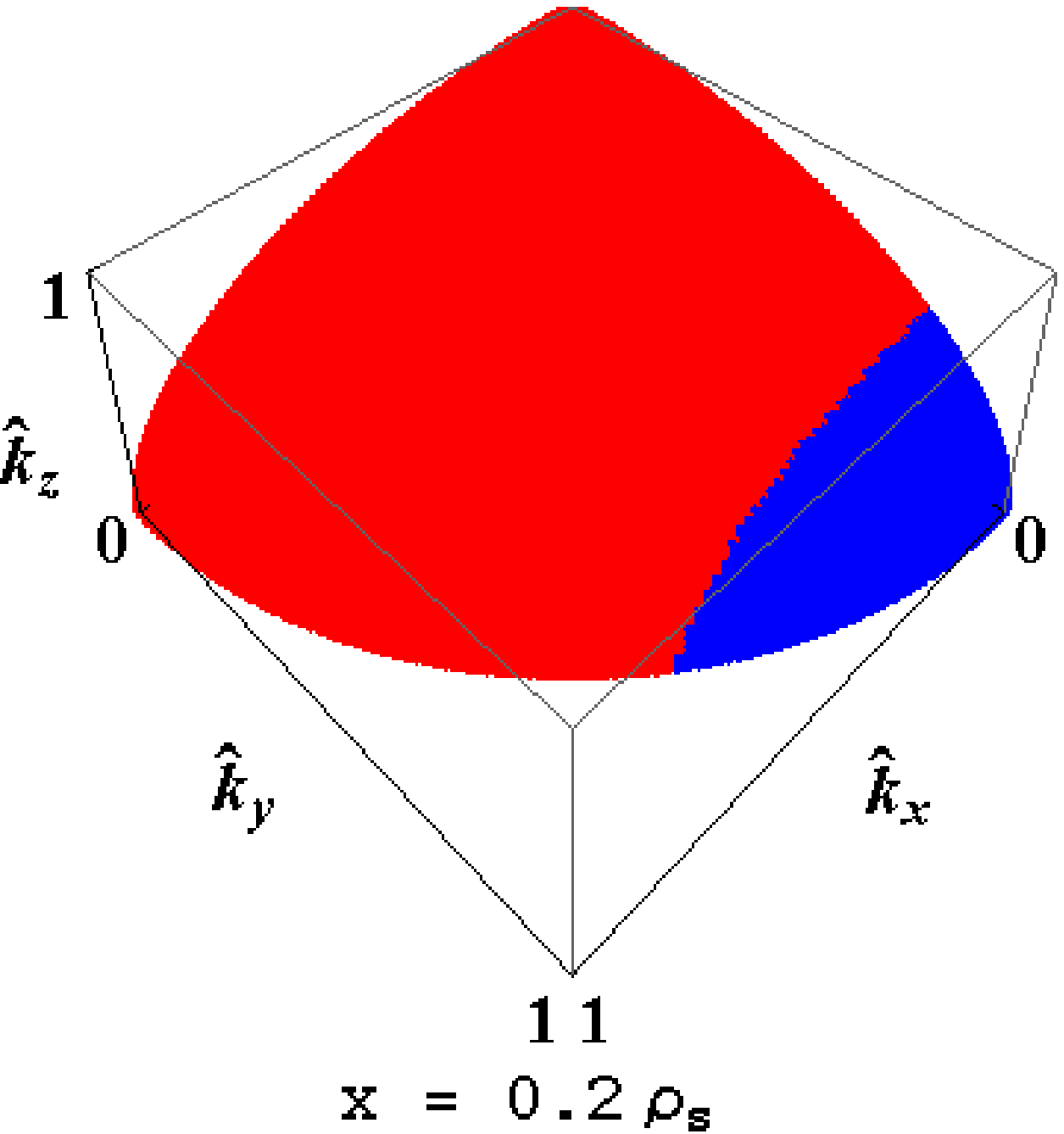,width=1.80in}
\epsfig{file=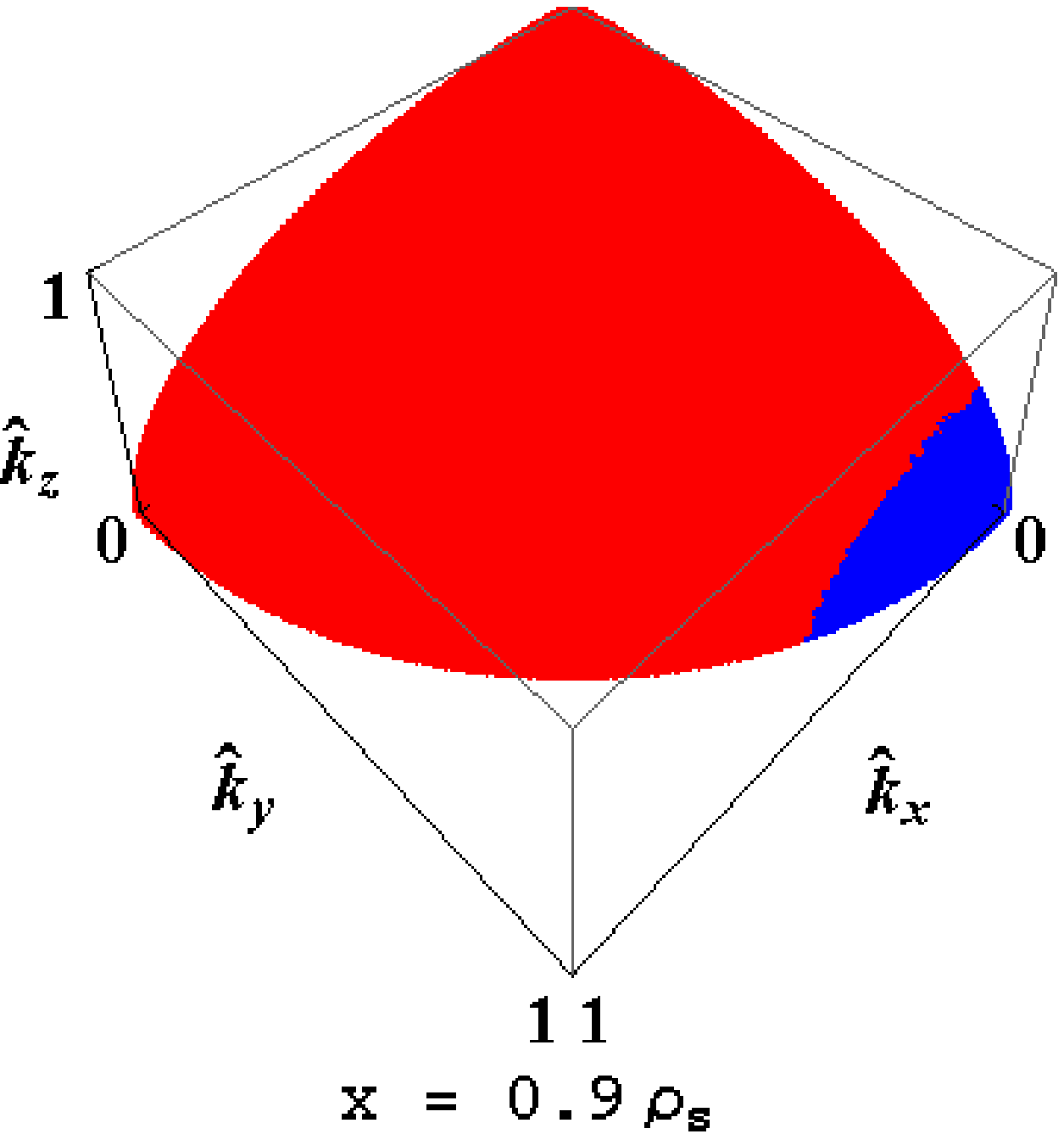,width=1.80in} \vspace{5mm}\\
\epsfig{file=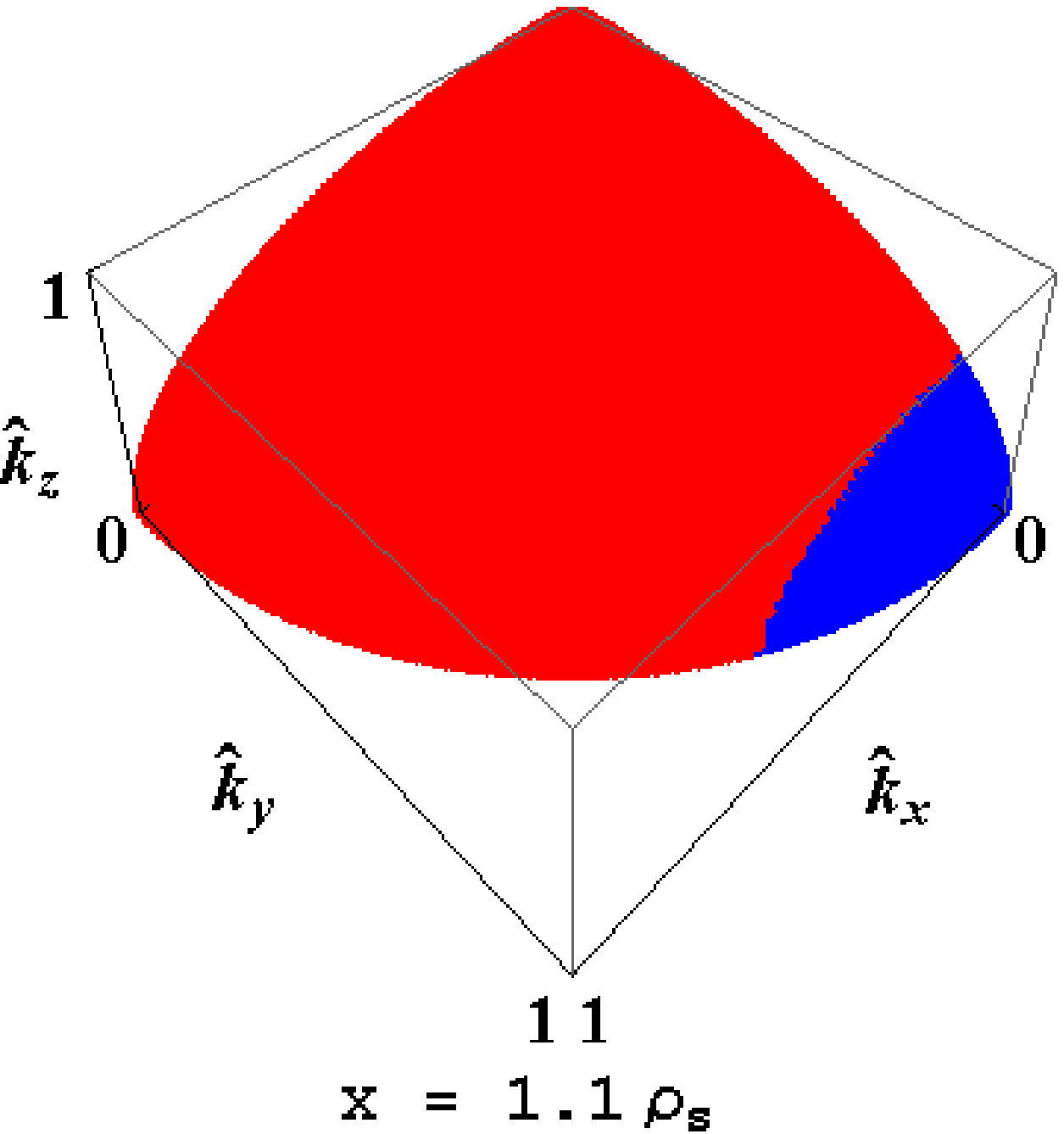,width=1.80in}
\epsfig{file=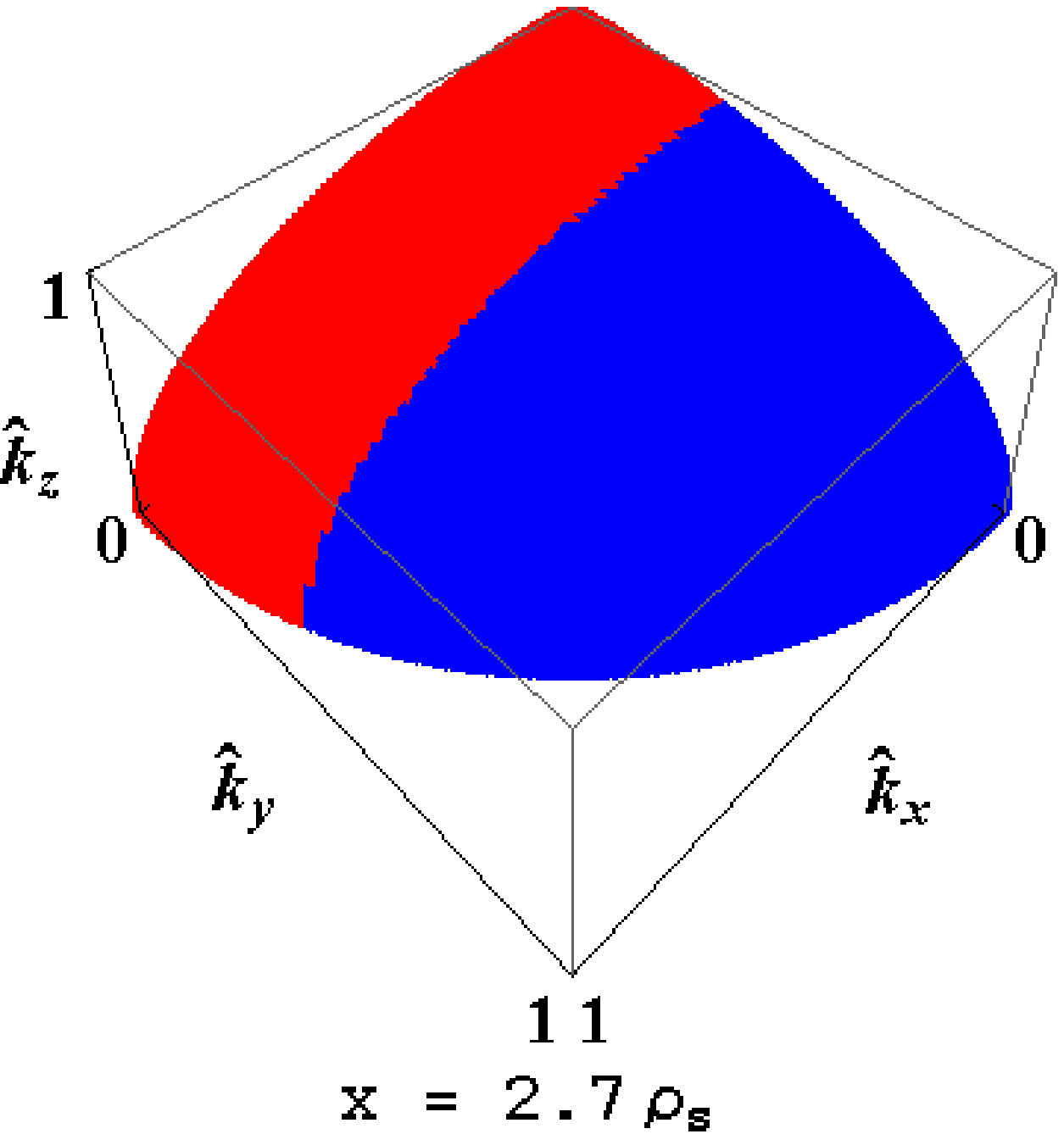,width=1.80in}
\epsfig{file=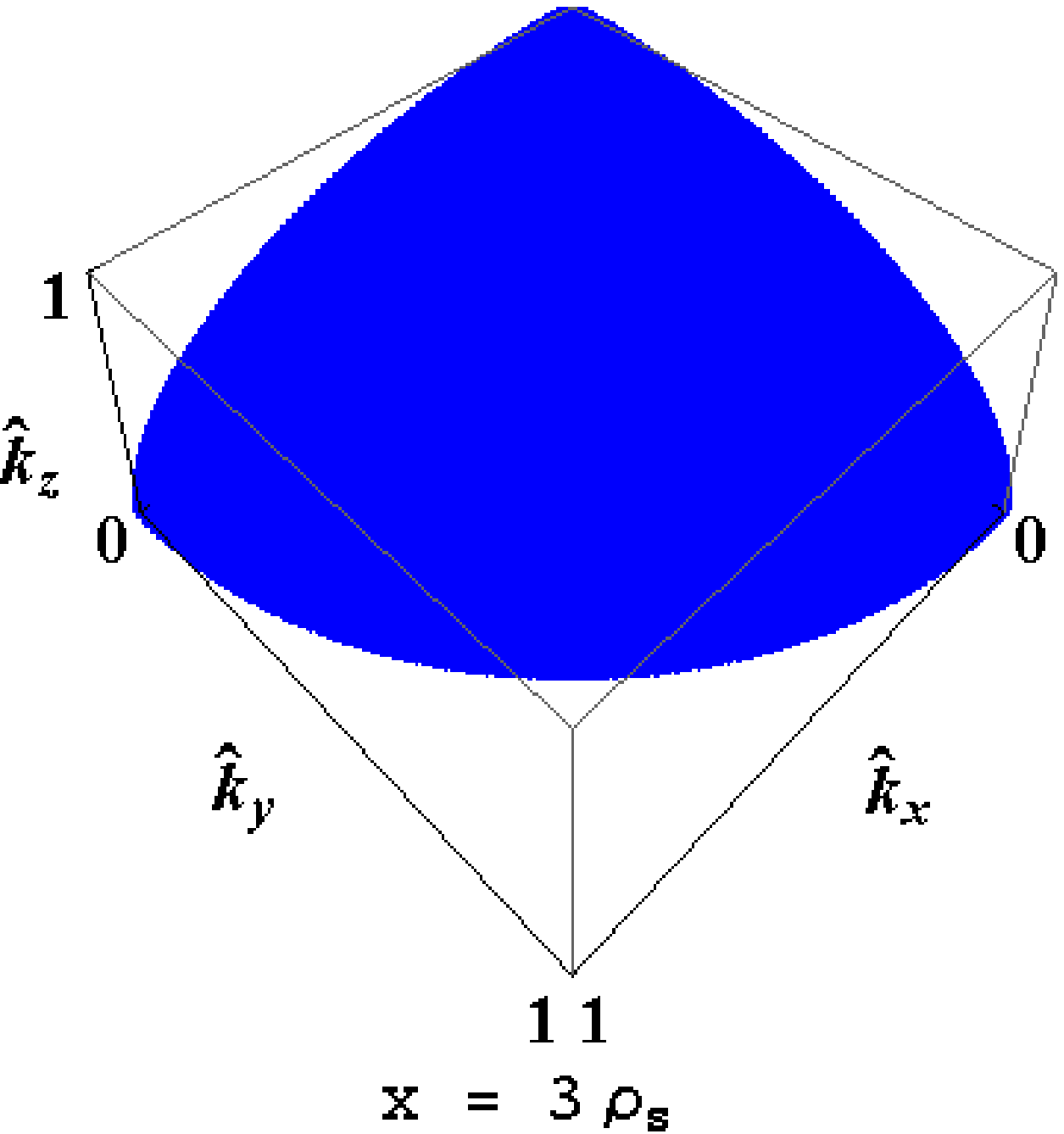,width=1.80in} \caption{Maps illustrating the
directions of $\#k = k \le \hat{k}_x, \hat{k}_y, \hat{k}_z \ri$ for
which $\mbox{Im} \lec k \ric = 0$ (blue) and $\mbox{Im} \lec k \ric
\neq 0$ (red),
 at locations along the $x$ axis with  $x \in \lec
0.1, 0.2, 0.9, 1.1, 2.7, 3.0 \ric \rho_s$. } \label{fig1}
\end{figure}

In general, $k$ can be complex--valued with non--zero imaginary
part. However, regimes where $\mbox{Im} \lec  k \ric \neq 0$
correspond to evanescent waves and are therefore excluded from our
study of ray trajectories. It is illuminating to characterize the
phase space corresponding to $\mbox{Im} \lec  k \ric \neq 0$. The
discriminant term in eq.~\r{kroot}, namely,
\begin{eqnarray}
b^2 - 4 a c &=& 4 \Big( M^2 \rho^2 \sin^2 \theta \, \sin^2 \le
\sigma - \phi \ri \nonumber \\ && -  \lec A^2 \cos^2 \theta + \les
A^2 \cos^2 \le \sigma - \phi \ri + \rho^2 \sin^2  \le \sigma - \phi
\ri \ris \sin^2 \theta
 \ric \le M^2 - A^2 \ri \Big), \l{disc}
\end{eqnarray}
can only be negative--valued when $A^2 -M^2 < 0$. Therefore, the regime
where evanescent waves arise coincides exactly with the regime where
the spacetime of a cosmic spinning string can support closed
timelike curves \c{css,Slobodov}. This regime is illustrated in
Fig.~\ref{fig1}, wherein the directions of $\#k$  for which $\mbox{Im}
\lec  k \ric \neq 0$ are depicted
 at locations along the $x$ axis. For reasons
of symmetry, only the directions in one octant of the unit sphere
need be shown. We see that the $\mbox{Im} \lec  k \ric \neq 0$
regime is restricted to the range $0.1 < \le \rho / \rho_s \ri < 3$.
Within this range, $\mbox{Im} \lec  k \ric \neq 0$ for most $\#k$
directions,  but  for $\#k$ directed  tangentially  $k$ remains
real--valued. It is clear from eq. \r{disc} that for a non--spinning
string (i.e., $M= 0$), $k$ is real--valued for all $\#k$ directions,
at all locations.

\section{Ray trajectories}

The scalar quantity $\mathcal{H} $ introduced in eq.~\r{disp} serves as
a convenient Hamiltonian function for our ray-tracing study.
 The ray trajectories are parameterized in terms of $\tau$ via  $\#r
(\tau)$, and the relative wavevector is likewise parameterized as $\#k
(\tau)$. The  coupled differential equations \c{Kline, Sluij2}
\begin{equation}
\left. \begin{array}{l} \displaystyle{
 \frac{d \#r}{d \tau} =  \nabla_{\#k}
 \mathcal{H}} \vspace{10pt} \\
\displaystyle{\frac{d \#k}{d \tau} = -  \nabla_{\#r}
 \mathcal{H}}
\end{array} \right\} \l{odes}
\end{equation}
 govern the trajectories of the light rays. Herein  the
shorthand $\nabla_{\#v} \equiv \le
\partial/ \partial v_x, \, \partial/ \partial v_y, \, \partial/ \partial
v_z \ri$ for $\#v = \le v_x, v_y, v_z \ri$ is adopted.

Before proceeding, we   confirm that the direction of  a ray
trajectory, i.e., $  \nabla_{\#k}  \mathcal{H} $, coincides with
the direction of  energy flux. We
 know from earlier studies with a general  Tamm medium that the
 time--averaged Poynting vector may be expressed as \c{MLS_NJP}
 \begin{equation}
\langle \, \#P \, \rangle_t = \beta \, \=\gamma (x, y) \cdot \#p\,,
 \end{equation}
where the scalar $\beta$ is positive--valued provided that $\=\gamma
(x, y)$ is either positive-- or negative--definite. From
eq.~\r{matrix_gamma}, the eigenvalues of $\=\gamma (x, y)$ are $\lec
A/\rho, A/\rho, \rho/A \ric$; hence $\beta > 0$. By direct vector
differentiation of the definition of  $\mathcal{H} $ provided in
eq.~\r{disp}, we find $ \nabla_{\#k}  \mathcal{H} = 2\, \=\gamma (x,
y) \cdot \#p$ . Therefore, the ray trajectories extracted from
eqs.~\r{odes} are indeed parallel to the direction of energy flux.

 Upon the specification
of appropriate initial conditions $\#r(0)$ and $\#k(0)$, the system
\r{odes} can be solved for $\#r( \tau)$~---~and $\#k (\tau)$~---~by
standard numerical methods, such as the Runge--Kutta method
\c{Sluij2}. For  a spinning string, two examples of  ray
trajectories are plotted in Fig.~\ref{fig2}. In the first example,
the rays start at locations in the $xy$ plane, with $\#k (0)$
directed along $\le -1, 0, 0 \ri$. In this case the ray trajectories
remain in the $xy$ plane. In the second example,  $\#k (0)$ is
directed along $\le -1, 0, -1 \ri$ and the ray trajectories are not
restricted to one plane. In both examples, the ray trajectories
skirt around the string's boundary,  never crossing it.

Let us now consider further whether it is  possible for ray
trajectories to cross the string's boundary. In Fig.~\ref{fig3},
trajectories are presented for rays which start close to the string
boundary, at $\#r(0) = \le 1.1, 0, 0 \ri \rho_s$ and $ \le 0.9, 0,
0 \ri \rho_s$. Regardless of the direction of $\#k (0)$, we find
that rays which start outside the string's boundary  remain outside,
and those which start inside the string's boundary remain inside. We
therefore conclude that rays cannot cross the string's boundary, in
either direction. Note that a ray  starting right at the string's
boundary cannot be considered, as the geometric-optics approximation
is not valid at this location.

\section{Invisibility}

\begin{figure}[!ht]
\centering \psfull \epsfig{file=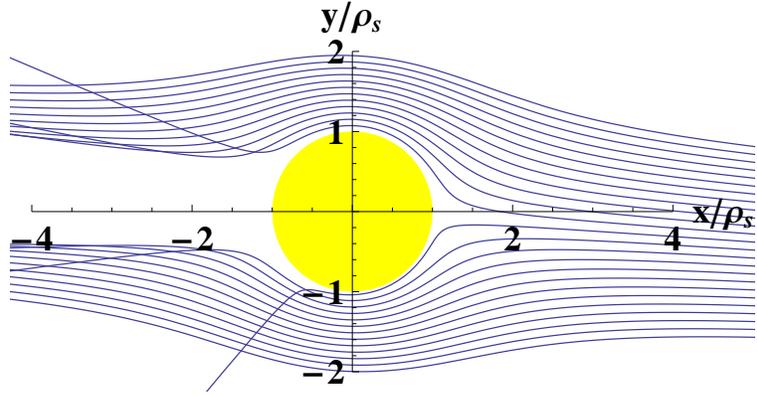,width=3.9in}
 \vspace{10mm}\\
\epsfig{file=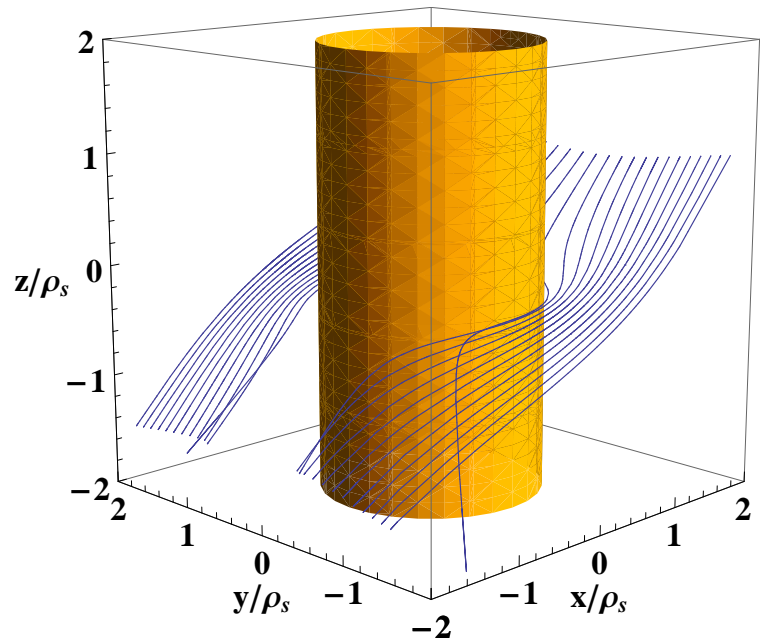,width=3.9in}
 \caption{2D and 3D examples of
ray trajectories for a spinning string. Top: Rays start at equally
spaced locations along the line $\#x (0) = \le 40, \nu, 0 \ri
\rho_s$ with $-3 < \nu < -1 $, and $\#k (0)$ directed along $\le -1,
0, 0 \ri $. Bottom: Rays start at equally spaced locations along the
line $\#x (0) = \le 40, \nu, 39 \ri \rho_s$ with $-3.9 < \nu < -1.9
$, and $\#k (0)$  directed along $\le -1, 0, -1 \ri$.} \label{fig2}
\end{figure}

\begin{figure}[!ht]
\centering \psfull \epsfig{file=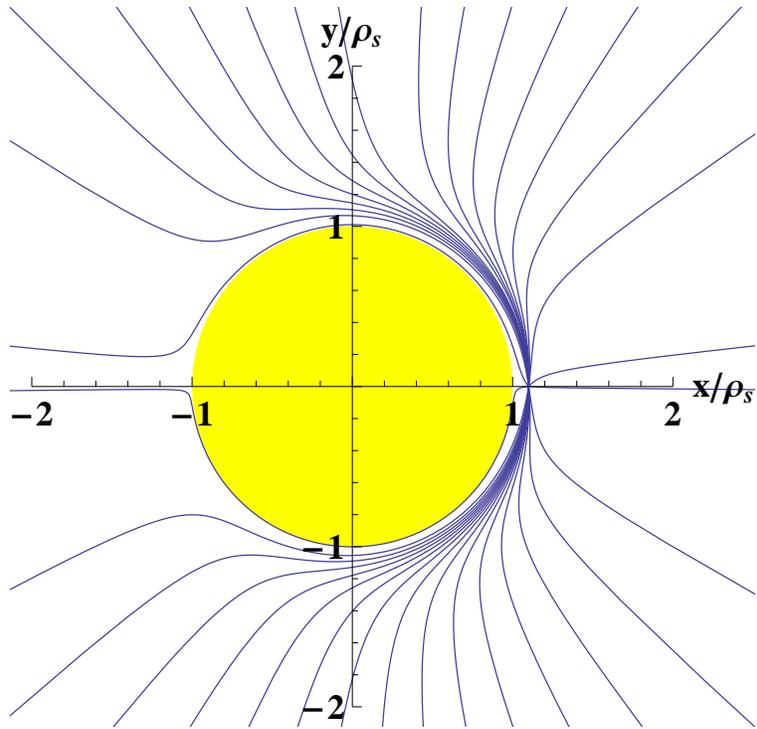,width=3.9in}
 \vspace{10mm}\\
\epsfig{file=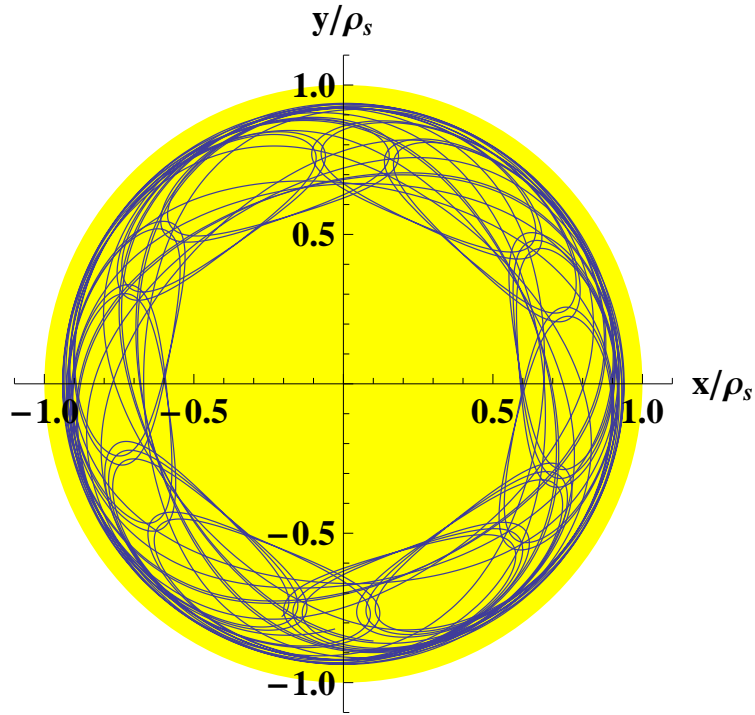,width=3.9in}
 \caption{Examples of 2D ray trajectories that start at locations close to the spinning string's boundary: $\#r (0) =  \le 1.1, 0, 0 \ri \rho_s$
(top)
  and $ \le 0.9, 0, 0 \ri
\rho_s$ (bottom). In both cases $\#k (0)$ is oriented at equally
spaced angular directions in the $xy$ plane. Note that for both
locations  approximately 70\% of the possible  orientations of $\#k$
correspond to evanescent waves and are therefore not represented
here.} \label{fig3}
\end{figure}

The ray trajectories presented in Fig.~\ref{fig2} are reminiscent of
those associated with metamaterials which are currently being
investigated for cloaking applications
\c{Greenleaf,Shin,Liu,Hashemi}. That is, the region
inside the spinning string's boundary is not visible to a distant
observer. However, this self-cloaking cloaking effect is not perfect as the ray
trajectories reaching a distant observer are distorted somewhat by
the string.

\begin{figure}[!ht]
\centering \psfull \epsfig{file=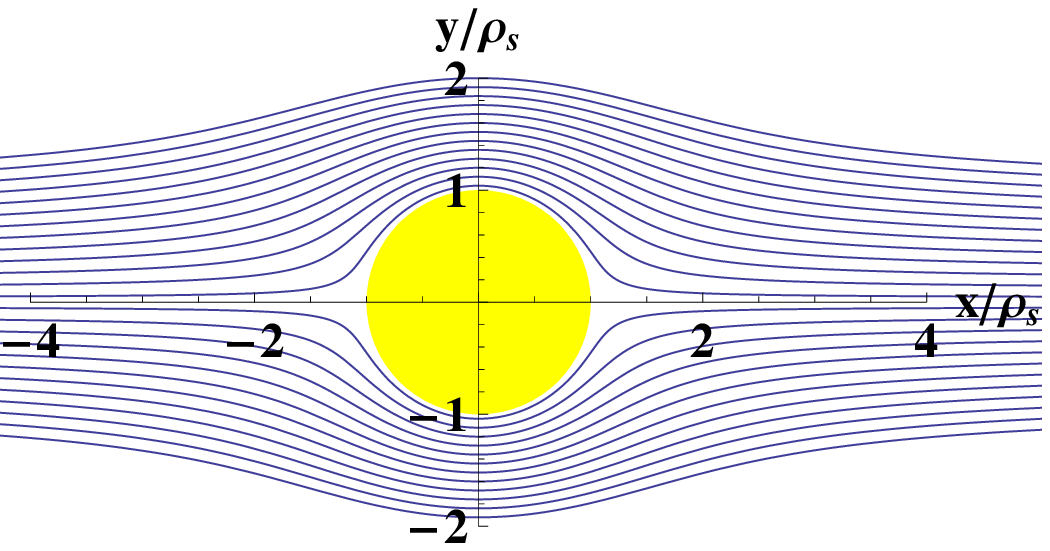,width=3.9in}
 \vspace{10mm}\\
\epsfig{file=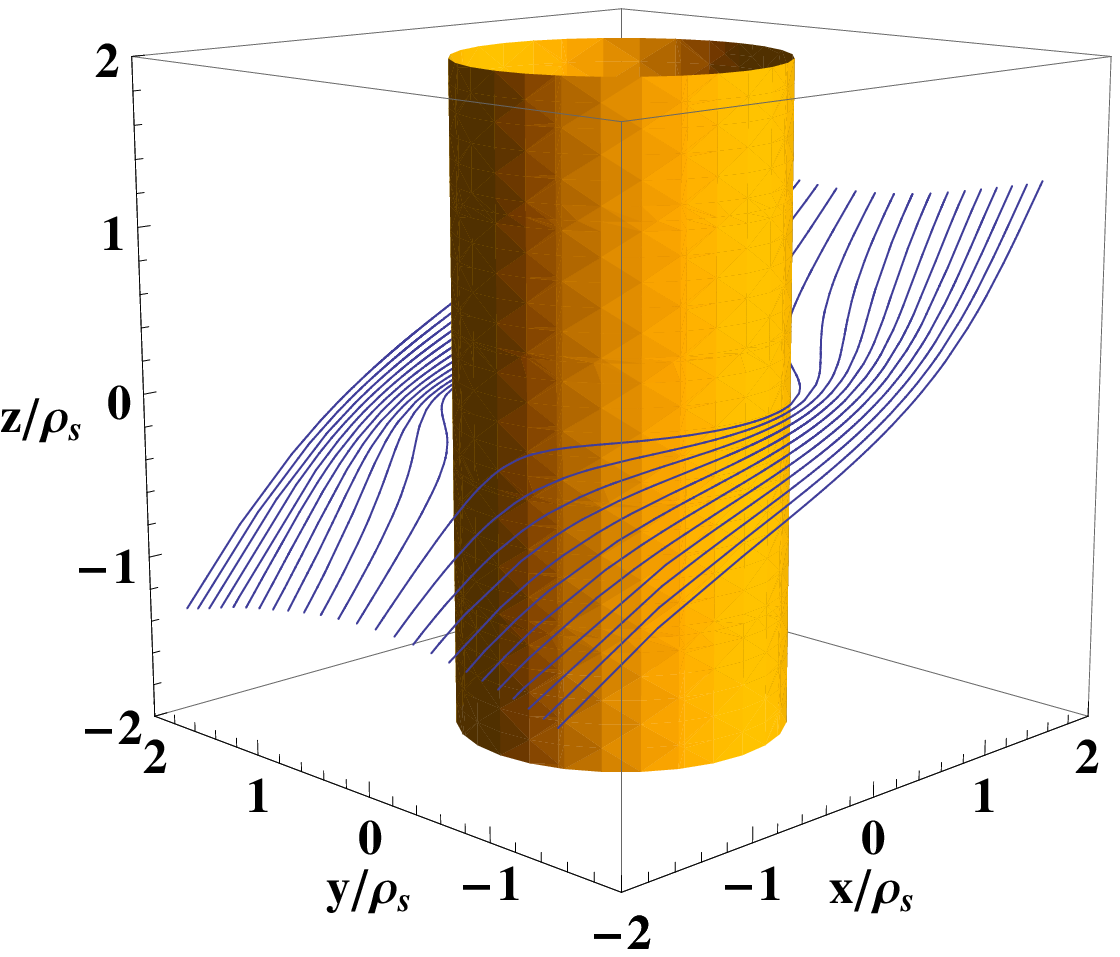,width=3.9in}
 \caption{As Fig.~\ref{fig2} except that the string is not spinning
and $-1 < \nu < 1$.} \label{fig4}
\end{figure}

In order to explore this matter further, let us consider a
non--spinning string. The ray trajectories for a non--spinning
string~---~corresponding to those presented in Fig.~\ref{fig2} for a
spinning string~---~are presented in Fig.~\ref{fig4}. We see that in
a radial plane,  the non--spinning string acts as an excellent
cloak for itself: the interior of the string is hidden to a distant observer
and there is minimal distortion to the ray trajectories reaching an
observer more than a few radiuses away from the  string. For rays
traversing  the string's neighborhood  at an oblique angle to the
$z$ axis, there is a small  degree of ray distortion  close to the
string~---~but this distortion diminishes as the distance from the
string increases.

\section{Closing remarks}

 The Tamm medium provides a
convenient setting for simulating the passage of light through the
spacetime associated with
 a cosmic spinning string. Based on a geometric-optics study,
 we have found that
 \begin{itemize}
\item ray trajectories do not cross the string's boundary; i.e., rays
which start  outside the string remain outside, whereas rays which
start inside the string's boundary remain trapped there;
\item the regions of the  spinning string's spacetime which support
closed timelike curves correspond to regions in the phase space of
the Tamm medium which support evanescent waves; and
\item a spinning string acts as an imperfect cloak for itself, while a
non--spinning string cloaks itself almost perfectly.
\end{itemize}
These findings may well have far--reaching astronomical
consequences, especially for those attempting to observe  cosmic
strings optically.

\vspace{10mm}

\noindent {\bf Acknowledgments:} THA is supported by an EPSRC (UK)
Vacation Bursary. TGM is supported by a Royal Academy of
Engineering/Leverhulme Trust Senior Research Fellowship. AL thanks
the Charles Godfrey Binder Endowment at Penn State for partial
financial support of his research activities.


\newpage

\newpage

\newpage

\newpage

\begin{thebibliography}{99}



\bibitem{de_Sitter_metamaterial}
M. Li, R.--X. Miao, Y. Pang,
Phys. Lett. B 689 (2010) 55. 


\bibitem{Black_hole_metamaterial}
Q. Cheng, T.J. Cui, W.X. Jiang, B.G. Cai,
New J. Phys. 12 (2010) 063006.


\bibitem{css} T.G. Mackay, A. Lakhtakia,
Phys. Lett. A 374 (2010) 2305. 


\bibitem{ML_PRB}
T.G. Mackay, A. Lakhtakia,
Phys. Rev. B  79 (2009)  235121.





\bibitem{MLS_NJP_Kerr}
T.G. Mackay, A. Lakhtakia,  S. Setiawan,
 New J. Phys. 7 (2005) 171.

\bibitem{Kom}
S.S. Komissarov,  J. Kor. Phys. Soc.  54 (2009) 2503.

\bibitem{Sharif}
M. Sharif, U. Sheikh, J. Kor. Phys. Soc.  55 (2009) 1677.

\bibitem{Kerr_Newman}
B.M. Ross, T.G. Mackay, A. Lakhtakia,
 Optik
 121 (2010) 401.

\bibitem{Hossain}
M. Hossain, M. Khayrul Hasan, Int. J. Theor. Phys.  48 (2009) 3007.


\bibitem{Tamm}
I.E. Tamm, Zhurnal Russkogo Fiziko-Khimicheskogo Obshchestva, Otdel
Fizicheskii (J. Russ. Phys.-Chem. Soc., Phys. Section)  56 (1924)
248.


\bibitem{Skrotskii}
G.V. Skrotskii,
 Soviet Phys.--Dokl.  2 (1957) 226.


\bibitem{Plebanski}
J. Pl\'{e}banski,
Phys. Rev.  118 (1960) 1396.


 \bibitem{SS}
W. Schleich,  M.O. Scully, in \emph{New Trends in Atomic Physics},
eds. G. Grynberg,  R. Stora  (Elsevier Science Publishers,
Amsterdam, Holland) 1984, p.~995.


\bibitem{Huebner}
U. Huebner, J. Petschulat, E. Pshenay-Severin, A. Chipouline, T.~Pertsch,
C.~Rockstuhl, F.~Lederer,
Microelectron. Eng. 86 (2009) 1138.

\bibitem{He}
P. He, J. Gao, Y. Chen, P. V. Parimi, C. Vittoria, V. G. Harris,
J. Phys. D: Appl. Phys. 42 (2009) 155005.

\bibitem{Hindmarsh}
M.B. Hindmarsh, T.W.B. Kibble, Rep. Prog. Phys. 58 (1995) 477.

\bibitem{Vilenkin}
A. Vilenkin, E.P.S. Shellard, \emph{Cosmological Strings and Other
Topological Defects} (Cambridge University Press, Cambridge, UK)
2000.

\bibitem{Greenleaf}
A. Greenleaf, Y. Kurylev, M. Lassas, G. Uhlmann,
SIAM Rev. 51 (2009) 3.

\bibitem{Jensen_Soleng}
B. Jensen, H.H. Soleng, Phys. Rev. D  45 (1992) 3528.

\bibitem{vanB}
J. van Bladel \emph{Electromagnetic Fields}
(Hemisphere, Washington, DC, USA) 1985, pp.~269--274.





\bibitem{LM1} A. Lakhtakia,  T.G. Mackay,
J. Phys. A: Math.  Gen.
  37 (2004) L505. Erratum  37 (2004) 12093.



\bibitem{LMcpl}
 A. Lakhtakia,  T.G. Mackay,
Chin. Phys. Lett.   23 (2006) 832.


\bibitem{Slobodov}
S. Slobodov, Found. Phys.  38 (2008) 1082.



\bibitem{Kline}
M. Kline, I.W. Kay, \emph{Electromagnetic Theory and Geometric
Optics} (Interscience, New York, NY, USA) 1965, pp.~109--112.


\bibitem{Sluij2}
M. Sluijter, D.K. de Boer,  H.P. Urbach, 
 J. Opt. Soc. Amer. A  26 (2009) 317. 

\bibitem{MLS_NJP}
 T.G. Mackay, A. Lakhtakia, S. Setiawan,
New J. Phys.
  7 (2005) 75.




\bibitem{Shin}
J. Shin, J.-T. Shen, S. Fan,
Phys. Rev. Lett. 102 (2009)  093903.


\bibitem{Liu}
R. Liu, C. Ji, J.J. Mock, J.Y. Chin, T.J. Cui, D.R. Smith, 
Science 323 (2009) 366. 

\bibitem{Hashemi}
H. Hashemi, B. Zhang, J.D. Joannopoulos,  S.G. Johnson,
Phys. Rev. Lett.  104 (2010) 253903.

\end{thebibliography}
\end{document}